\documentstyle[12pt]{article}

\def\a{\begin{eqnarray}}
\def\b{\end{eqnarray}}
\def\0{\nonumber}
\input amssym.def
\font\teneusm=eusm10                    
\font\seveneusm=eusm7                   
\font\fiveeusm=eusm5                    
\newfam\eusmfam
\textfont\eusmfam=\teneusm
\scriptfont\eusmfam=\seveneusm
\scriptscriptfont\eusmfam=\fiveeusm

\renewcommand{\theequation}{\thesection.\arabic{equation}}

\newlength{\extraspace}
\newlength{\extraspaces}
\newcounter{dummy}

\newcommand{\ai}{
\addtocounter{equation}{1}
\setcounter{dummy}{\value{equation}}
\setcounter{equation}{0}
\renewcommand{\theequation}{\thesection.\arabic{dummy}\alph{equation}}
\begin{eqnarray}
\addtolength{\abovedisplayskip}{\extraspaces}
\addtolength{\belowdisplayskip}{\extraspaces}
\addtolength{\abovedisplayshortskip}{\extraspace}
\addtolength{\belowdisplayshortskip}{\extraspace}}
\newcommand{\bj}{
\end{eqnarray}
\setcounter{equation}{\value{dummy}}
\renewcommand{\theequation}{\thesection.\arabic{equation}}}

\input amssym
\newcommand{\be}{\begin{equation}}
\newcommand{\ee}{\end{equation}}
\newcommand{\ba}{\begin{eqnarray}}
\newcommand{\ea}{\end{eqnarray}}
\newcommand{\ban}{\begin{eqnarray*}}
\newcommand{\ean}{\end{eqnarray*}}
\newcommand{\brr}{\begin{array}}
\newcommand{\err}{\end{array}}
\newcommand{\bc}{\begin{center}}
\newcommand{\ec}{\end{center}}

\def\D{{\cal D}}
\def\E{{\cal E}}
\def\G{{\cal G}}

\def\L{ \Lambda}

\def\S{{\cal S}}

\def\F{{\cal F}}
\def\l{\lambda}
\def\s{\sigma}
\def\al{\alpha}
\def\be{\beta}
\def\ga{\gamma}
\def\de{\delta}
\def\ep{\epsilon}
\def\o{\omega}
\def\co{\Omega}

\def\var{\varphi}
\def\z{\zeta}
\newcommand{\bea}{\begin{eqnarray}}
\newcommand{\eea}{\end{eqnarray}}
\newcommand{\bean}{\begin{eqnarray*}}
\newcommand{\eean}{\end{eqnarray*}}
\newcommand{\CC}{\Bbb C}

\newcommand{\ZZ}{\Bbb Z}

\newcommand{\del}{\partial}

\begin{document}

\begin{titlepage}

\begin{flushright}
CBPF--NF--050/98
\end{flushright}

\vskip0.5cm
\centerline{\LARGE $A_n^{(1)}$ Toda Solitons:}
\vskip0.5cm 
\centerline{\LARGE a Relation between
 Dressing transformations}
\vskip0.5cm 
\centerline{\LARGE and Vertex Operators\footnote{Talk given at the iV 
International Conference on Non Associative Lie Algebra and its Applications, 
University of S\~ao Paulo, July 19--20, 1998, S\~ao Paulo, Brazil}}

\vskip1.5cm
\centerline{\large   H. Belich
\footnote{E--mail address belich@cbpfsu1.cat.cbpf.br}} 
\vskip0.5cm
\centerline{\large and }
\vskip0.5cm
\centerline{\large R. Paunov
\footnote{E--mail address paunov@cbpfsu1.cat.cbpf.br}}
\vskip0.5cm
\centerline{Centro Brasileiro de Pesquisas Fisicas }
\centerline{Rua Dr. Xavier Sigaud 150, Rio de Janeiro, Brazil}

\vskip2.5cm
\abstract{ Affine Toda equations based on simple Lie algebras
arise by imposing zero curvature condition on a Lax connection 
which belongs to the corresponding loop Lie  algebra in the 
principal gradation. In the particular case of $A_n^{(1)}$
 Toda models, we exploit the symmetry of the underlying 
linear problem to calculate the dressing group element 
which generates arbitrary $N$-soliton solution from 
the vacuum. Starting from this result we recover the vertex 
operator representation of the soliton tau functions.}

\end{titlepage}

\section{ Introduction}

\setcounter{equation}{0}
\setcounter{footnote}{0}

   Integrable nonlinear evolution equations play important
 role in various topics of the mathematics and 
particle physics. There is a huge class of integrable
equations which admit soliton solutions \cite{Dic}. The last are exact localizable solutions of the equations of motion with finite values of the physical observables: momentum, energy, etc. \cite{Raj}. Generically soliton solutions correspond to topologically nontrivial field configurations. Within the quantum field theory, the solitons are interpreted  as nonperturbative particles which appear in the spectrum of the underlying model.

   Integrable evolution equations are distinguished by the existence of zero curvature or Lax representation. Taking into account the Lax representation of the equations of motion, one can look for {\it gauge} transformations which preserve the form of the components of the Lax connection. It is clear that such kind of transformations, called dressing transformations, are symmetries of the  underlying integrable model. The study of the dressing group which is a symmetry of the soliton (integrable) equation has been initiated by the Kyoto group \cite{jap}. In the last reference the dressing symmetry was considered on the example of the Kadomtsev-Petiashvili (KP) hierarchy. The group of dressing transformations admits a particularly simple and transparent expression when one introduces the Hirota tau functions. The last provide a bridge between integrable models and the representation theory of the affine Lie algebras \cite{Kac}. It has been shown by Semenov-Tian-Shansky \cite{STS} that the dressing group is a Poisson Lie group.

   The application of the dressing group to the Toda field theories in $1+1$ dimensions is due to Babelon and Bernard \cite{fr}. In this reference it was also argued that the dressing symmetry is a semiclassical analog of the quantum symmetry of the integrable model.

   The $N$ soliton solutions of the affine Toda models based on an arbitrary simple Lie algebra are obtained in \cite{DT}. In particular the group-theoretical tau functions related to the fundamental representations of the affine Lie algebra are given by
\ai
&& \hskip -2cm \frac{\tau_{\L}(\Phi)}{\tau_{\L}(\Phi_0)}=
e^{\frac{\z_0-\z}{n+1}}\prod_{i=1}^N Y_i \cdot
<\L|(1+X_1F^{r_1}(\mu_1))\ldots (1+ X_N F^{r_N}(\mu_N))|\L>
\label{1.1a}
\b
where $X_i, i=1,...,N$ are numerical factors depending exponentially on the light cone coordinates, $F^{r_i}$ are elements of the affine Lie algebra which diagonalize the adjoint action if the principal Heisenberg subalgebra \cite{Kac, DT} and $|\L>$ is the highest weight vector of a fundamental representation of the affine Lie algebra of highest weight $\L$. Alternative expression for the tau functions is provided by using the dressing symmetry
\a
\frac{\tau_{\L}(\Phi)}{\tau_{\L}(\Phi_0)} = 
<\L|\hat{g}_-^{-1}( x^+, x^-)\, \hat{g}_+( x^+, x^-)|\L>
\label{1.1b}
\bj
where $\hat{g}_-$ and $\hat{g}_+$ are triangular elements of the affine group.

   In \cite{BB} it has been shown that the representations (\ref {1.1a}) and (\ref {1.1b})
 are equivalent  for the  the sine-Gordon $N$--solitons. Our purpose is to extend this result for arbitrary $A_n^{(1)}$ Toda models. General analysis of the dressing symmetry of integrable hierarchies which admit vacuum solution has been done in \cite{Fer}.

   In the present talk we give a summary of  \cite{Bel}. It is organized as follows:
   In Sec.1 we discuss the $N$-soliton solutions of the $A_n^{(1)}$ Toda equations and the related dressing problem. An explicit expression for the dressing transformation which creates solitons from the vacuum is found. The main result of Sec.1 is the Proposition 2. It states that the above mentioned dressing transformation factorizes into "monosoliton" factors. In Sec.3 we exploit the free field representations of the affine Lie algebra $A_n^{(1)}$ to demonstrate the equivalence between (\ref {1.1a}) and (\ref {1.1b}). Our approach is based on the observation that the solution of the dressing problem on the affine group differs from those in the loop group by a factor which is in the center of the affine group. Sec.4 is reserved for concluding remarks and discussion of possible further developments.

\section{ $N$-solitons and the solution of the dressing problem in  the affine group}

\setcounter{equation}{0}
\setcounter{footnote}{0}

 The $A_n^{(1)}$ Toda equations in $1+1$ dimensions \cite{OT}
\a
&&\del_+\del_-\var_i=m^2(e^{\var_i-\var_{i+1}}
-e^{\var_{i-1}-\var_i}),\0\\
& &\del_{\pm}=\frac{\del}{\del x^{\pm}},~~~
i \in  {\ZZ}_{n+1}\label{2.1}
\b
are equivalent to the zero curvature condition
\ai
&&F_{+-}=\del_+A_--\del_-A_++\left[A_+, A_-\right]=0\label{2.2a}
\b
of a connection which belongs to the loop Lie algebra $\widetilde{sl}(n+1)=sl(n+1)\otimes \CC [\lambda,{\lambda}^{-1}]$\\
\a
A_\pm&=& \pm \del_\pm \Phi + m\l^{\pm 1}  e^{\pm ad\Phi}\E_{\pm}\0\\
 \del_{\pm}&=&\frac{\del}{\del x^\pm}
\label{2.2b}
\b
where $\Phi$ is in the Cartan subalgebra of $sl(n+1)=A_n$
\a
& &\Phi=\frac{1}{2}\sum_{i}\var_i |i><i|
\label{2.2c}
\b
and
\a
\E_{\pm}= \sum_{k\in {\ZZ}_{n+1}} |k><k \pm 1|. \label{2.2d}
\bj

 The adjoint action of the element $S\in SL(n+1)$
\ai
S&=& \o^{\frac{n}{2}} \sum_{k\in {\ZZ}_{n+1}} \o^{1-k} |k><k|~,~
 \o=e^{\frac{2\pi i}{n+1}}\label{2.3a}
\b
defines an inner automorphism $\s$ of order $n+1$ of
the algebra $A_n$
\a
\s(X)=SXS^{-1}, ~~~\s^{n+1}=1.\label{2.3b}
\bj
Applying $\s$ on the components
 of the Lax connection (\ref{2.2b}) we conclude that
\a
A_{\pm}(x,\o\l)=\s(A_{\pm}(x,\l))=SA_{\pm}(x,\l)S^{-1}\label{2.4}
\b
Since  (\ref{2.2b}) is a flat connection,  there is a covariantly constant vector $w(x,\l)$:
\a
D_{\pm}w(x,\l)= (\del_{\pm}+A_{\pm})w(x,\l)=0
\label{2.5}
\b
Taking into account (\ref{2.4}), one concludes that
$\S^r(w)(x,\l)=S^r w(x,\o^{-r}\l)$ for any
$r \in \ZZ_{n+1}$ is covariantly  constant also. This observation allows us to 
construct a matrix solution of the linear system (\ref{2.5})
\a
W(x,\l)&=&\parallel w(x,\l),
\o^{\frac{n}{2}}(\S^{-1}w)(x,\l)
\ldots 
\o^{\frac{n^2}{2}}(\S^{-n}w)(x,\l)\parallel
\label{2.6}
\b
To obtain $N$--soliton solutions of (\ref{2.5}) we introduce
 the expansion
\a
& &\hat{w}(x,\l)=\sum_{j=0}^N \l^jw^{(j)}(x)
e(x,-\l)\0\\
& &w^{(j)}=\sum_{k\in {\ZZ}_{n+1}}w^{(j)}_k |k>\0\\
& &e(x,\l)=exp \{m(\l x^++\frac{x^-}{\l})\}
\label{2.7}
\b
where $N$ is an integer which will be identified with 
the number of solitons and $w^{(j)}(x)$,  $j~=~1,...,N$ are 
$\l$--independent $n+1$--dimensional vectors. We shall need the following 

${\bf Proposition\,\, 1}$ \cite{Date} {\it Let} $w^{(N)}=\sum_{k\in {\ZZ}_{n+1}}
|k>$ {\it in the expansion (\ref{2.7}). Suppose also that for each 
 $j=1,...,N$ there is an integer $r_j=1,\ldots\,, n$ and complex numbers $\mu_j$,
$\mu_i^{n+1}\neq \mu_j^{n+1}$ such that }
\a
(\S^{-r_j}\hat{w})(x,\mu_j)&=&\o^{-\frac{r_jn}{2}}
c_j\hat{w}(x,\mu_j)\0\\
\label{2.8}
\b
{\it for arbitrary constants $c_j,\,\,j=1,\ldots ,\, N$.}
 
{\it Then $\hat{w}$ is a solution of the system}
\a
&& (\del_{\pm}+\hat{A}_{\pm}(x,\l))\hat{w}(x,\l)=0\0\\
&& \hat{A}_+(x,\l)=2\del_+\Phi(x)+m\l\E_+\0\\
&&\hat{A}_-(x,\l)=\frac{m}{\l}e^{-2ad\Phi(x)}\E_-
\label{2.9}
\b
{\it for certain traceless diagonal matrix $\Phi$.}

As a consequence from the above Proposition we see
that the entries of $\Phi$ satisfy the $A_n^{(1)}$ 
Toda equations (\ref{2.1}) Alternatively, the components
$\hat{w}_k$ of $\hat{w}$ (\ref{2.7}) can written as
\a
\hat{w}_k(x,\l)&=& \prod_{j=1}^N (\l+\ep_{kj}(x)) 
e(x,-\l),k\in {\ZZ}_{n+1} \0
\b
and from the Proposition $1$ it follows that the evolution
of the $N$-soliton system assumes the form
\ai
\prod_{l=1}^N \frac{\ep_{kl}+\o^{r_j}\mu_j}
{\ep_{kl}+\mu_j}&=&c_j \o^{r_j(1-k)}
\frac{e(\o^{r_j}\mu_j)}{e(\mu_j)}.\label{2.10a}
\b
The components of the Toda field are given by

\a
e^{-\var_k}&=& (-)^N \prod_{j=1}^N
\frac{\ep_{kj}}{\mu_j},~~~~k=1,\ldots , n+1
\label{2.10b}
\bj
Note that $\hat{w}(x,\l)$ (\ref{2.9}) is covariantly constant vector with
respect to the connection $\hat{D}_{\pm}=e^{-\Phi} D_{\pm} e^{\Phi}$ where
$D_{\pm}=\del_{\pm}+A_{\pm}$ is given by (\ref{2.2b}). Therefore, $w=e^{\Phi} \hat{w}$.

In accordance with the general definition, \cite{ jap, fr, Fer}, 
the dressing transformations are represented by loop group
 elements $g(x,\l) \in \widetilde{SL}(n+1)$ which act on the 
Lax connections (\ref{2.2b}) as gauge transformations 
 $A_{\pm} \rightarrow A_{\pm}^g$
\ai
A_{\pm}^g&=& - \del_{\pm} g g^{-1}+gA_{\pm}g^{-1},
\label{2.11a}
\b
such that the connection $A_{\pm}^g$ has the same form
 as the original one (\ref{2.2b}) with $\Phi \rightarrow \Phi^g$.
 Since under gauge transformations the curvature transforms
 as $F_{+-} \rightarrow g F_{+-} g^{-1}$, we see that the 
dressing group is a symmetry of the equations 
of motion. Equivalently, the action of the dressing group 
can be described in terms of the normalized transport 
matrix
\a
T(x,\l) \rightarrow T^g(x,\l)=g(x,\l)T(x,\l)g^{-1}(0,\l)\label{2.11b}
\b
where $T(x,\l)$ is a solution of the linear system
\a
\left( \del_{\pm} + A_{\pm} \right)T(x,\l)
 &=&0 \label{2.11c}
\bj
together with the initial value conditions
\a
&&T(0,\l)=T|_{x^+=x^-=0}=1\0
\b

The loop group $\widetilde{G}$ (in the present talk
 we consider $G= SL(n+1)$ only) is defined \cite{Ps} as a 
group of smooth maps of the unit circle $S^1 (|\l|=1)$
 into the Lie group $G$. It is clear that the loop group
 has two subgroups $\widetilde{G}_+$ and $\widetilde{G}_-$  : $\widetilde{G}_+\,(\widetilde{G}_-)$
 contains all the elements which admit analytic extension on $\l$
 inside (outside) the unit disc. In view of (\ref{2.2b}) $A_+, \, (A_-)$
 is holomorphic Lie valued function for $|\l|<1$, $(|\l|>1)$.
 Therefore, one can look for two solutions $g_{\pm} \in \widetilde{G}_{\pm}$
 of the dressing problem (\ref{2.11a}). Since $g_+$
 and $g_-$ produce the same result, from (\ref{2.11b}) it 
follows that they provide a solution of the factorization problem
\a
g_-^{-1}(x)g_+(x)=T(x)g_-^{-1}(0)g_+(0)T^{-1}(x)
\label{2.12}
\b
Note that  the transport matrix  related to the vacuum 
solution $\Phi=0$ (\ref{2.1}), (\ref{2.2c}) is $T_0(x,\l)= e^{-m(\l\E_+x^++\frac{x^-}{\l}\E_-)}$. In \cite{Bel} we have shown that there is a solution of (\ref{2.11a}) which interpolates between the vacuum
 and the $N$-soliton solution $\Phi$ (\ref{2.10a})-(\ref{2.10b})
\ai
& &g^{(N)}(\Phi,\{\mu\},\l)= e^{\Phi}
\Gamma^{(N)}(\Phi,\{\mu\},\l) \co^{-1}\0\\
& &\Gamma_{kl}^{(N)}(\Phi,\{\mu\},\l)=
 \o^{(k-1)(l-1)} \prod_{j=1}^N 
\frac{\l+\o^{1-l}\ep_{kj}(x)}{\l-\o^{1-l}\mu_j}\0\\
& &\co_{ij}= \o^{(i-1)(j-1)}~~~,~~~~\l \in ~~S^1
\label{2.13a}
\b
which belongs to loop group $\widetilde{SL}(n+1)$
and obeys the relation (c.f. (\ref{2.4}))
\a
g^{(N)}(\Phi,\{\mu\},\o \l)&=&
S g^{(N)}(\Phi,\{\mu\},\l) S^{-1}\label{2.13b}
\bj
To proceed, we have to introduce some Lie algebraic
background \cite{ Kac,DT}. First of all, note that
the automorphism $\s$ (\ref{2.3a}), (\ref{2.3b}) 
defines a $\ZZ_{n+1}$--gradation in the Lie algebra $\G = sl(n+1)$
\a
\G=\oplus_{k\in{\ZZ}_{n+1}}\G_k~~~,~~~
\s(\G_k)=\o^k\G_k\0\\
\label{2.14}
\b 
On the other hand, the elements
\ai
\E_{i}= \sum_{k\in {\ZZ}_{n+1}} |k><k+i|~~~,
~~~i \in {\ZZ}_{n+1} \backslash\{0\}   \label{2.15a}
\b
are mutually commuting and diagonalizable
\a
&&\co^{\pm 1} \E_k \co^{\mp 1}=\o^{\pm \frac{kn}{2}} 
S^{\pm k}\0\\
&&\E_1=\E_+~~~,~~~~~~\E_{-1}=\E_n=\E_{-}
\label{2.15b}
\bj
where $\co$ is defined by the last equation (\ref{2.13a}).
Therefore, the set of generators $\E_k$, $\E_k \in \G_k$
span a new Cartan subalgebra. To complete it to a base
of the Lie algebra we define the elements 
\ai
F^i=\co |i+1><i| \co^{-1}\quad , \quad i \in {\ZZ}_{n+1}
 \backslash \{0\}
\label{2.16a} 
\b   
 and consider their grade expansions
\a
F^{i}= \sum_{k\in {\ZZ}_{n+1}} F^{i}_k,
~~~~~~~\s(F^{i}_k)=\o^k F^{i}_k 
\label{2.16b}
\bj
The commutation relations in the new basis are
\ai
\left[F^r_i , F^s_l\right]= \frac {\o^{sk}-\o^{rl}}{n+1}F^{r+s}_{k+l} 
\label{2.17a}
\b
where
\a
F^r_k&=&\left\{\begin{array}
{ccc}
F^r_k & 
r=1, \ldots, n;& k\in {\ZZ}_{n+1}\\ \\
 \frac{1}{n+1}\E_{k}
& r=0 & k=1, \ldots, n 
\end{array} \right. 
\label{2.17b}
\b
In the above basis the invariant non degenerate
scalar product $(X,Y)=tr(X.Y)$ (the trace is taken
in the defining representation ) is given by 
\a
(F^r_k ,F^s_l)=\frac{\o^{sk}}{n+1} \delta^{(n+1)}_{k+l,0}
\label{2.17c}
\bj
$\delta^{(n+1)}_{i,j}$ is the Kronecker symbol on the
cyclic group.

To treat integrable evolution equations one has
to extend the classical Lie algebra by introducing a
spectral parameter $\l$ (c.f. (\ref{2.2b})). 
The Lax connection is in the loop Lie algebra 
$\tilde{\G}={\CC}[\l,\l^{-1}] \otimes \G$. $\tilde{\G}$
is spanned on the elements $X_n=\l^nX$. Generic loop algebra 
posses  a central extension \cite{ Kac, Ps} known as 
affine Lie algebra $\widehat{\G}=
\widetilde{\G} \oplus {\CC}\widehat{c} \oplus {\CC}\widehat{d}$
\ai
& &\left[X_k , Y_l\right]= \left[X,Y\right]_{k+l}
+\frac{k}{(n+1)}\hat{c}\, \delta_{k+l,0}\, (X,Y)\0\\
& &\left[\hat{d} , X_k\right]=k X_k~~~~,~~~~~~
\left[\hat{c} , \hat{\G}\right]= 0
\label{2.18a} 
\b 
The derivation $\widehat{d}=\l \frac{\del}{\del\l}$ 
defines a $\ZZ$-gradation 
\a
\tilde{\G}=\oplus_{k\in{\ZZ}}\tilde{\G}_k~~~,~~~~
\left[ \widehat{d}, \widehat{\G}_k \right]=k\widehat{\G}_k
\label{2.18b}
\bj
Note also that setting $\widehat{c}=0$ in (\ref{2.18a}), 
one recovers the commutation relations in the loop algebra.

The symmetry of the Lax connection (\ref{2.4})
suggests the following.

${\bf Definition}$: {\it The affine ( loop) Lie algebra $\widehat{sl}(n+1)\,\,\,(\widetilde{sl}(n+1))$
is in the principal gradation iff its elements 
$X(\l)=\sum_{l\in {\ZZ}}\l^lX_l$ satisfy the restriction} 
\a
X(\o \l)=\s X(\l)
\label{2.19}
\b
In view of the above definition, (\ref{2.4}) is in 
the loop algebra $(\tilde{sl}(n+1))$ in the principal gradation.
The affine algebra analog of the base (\ref{2.17b}) is
\ai
&&F^r_k,~~~~~~~r=1,\ldots ,\,n,\,\,\, k\in {\ZZ}\0\\
&&F^0_k,~~~~~~~ k\in {\ZZ} \ (n+1){\ZZ}
\label{2.20a}
\b
and the commutation relations are
\a
\left[F^r_k , F^s_l\right]=\frac{\o^{sk}-\o^{rl}}{n+1}F^{r+s}_{k+l}+ \frac{k\o^{rl}}{(n+1)^2}~\hat{c}~ \delta^{(n+1)}_{r+s,0}~\delta_{k+l,0}
\label{2.20b} 
\b
In the above notations and in what follows we will perform a slight abuse of notations: the lower index will be used to indicate both the cyclic  
$\ZZ_{n=1}$ charge (\ref{2.14}) as well as to label the $\ZZ$ gradation (\ref{2.18b}).
The elements $\E_k=(n+1)F^0_k~,~k\neq 0\,{\rm mod}~(n+1)$ form an infinite set of decoupled harmonic oscillators
\a
\left[\E_k , \E_l\right]= k~ \hat{c}\delta_{k+l,0}\,\,\,\,
~k\neq 0\,{\rm mod}(n+1)
\label{2.20c}
\bj
The subalgebra spanned on $\E_k$ is known as the
Heisenberg subalgebra in the principal gradation.
In order to formulate one of the main results of 
\cite{ Bel} we have to introduce some notations.
Let (\ref{2.10a}) be an arbitrary $N$-soliton solution.
To it we associate the diagonal traceless matrices
\ai
&&F_i=\frac{1}{2}\sum_{k\in{\ZZ}_{n+1}} f_{ki}|k><k|, \quad
P_i=\sum_{k\in{\ZZ}_{n+1}} p_{ki}|k><k|\0\\
&&K(F_i)=\sum_{k\in{\ZZ}_{n+1}}K_k(F_i)|k><k|, \quad i=1,\ldots\, ,N
\label{2.21.a}\\
&&K_k(F_i)-K_{k+1}(F_i)=\frac{f_{k~i}+f_{k+1~i}}{2}~~~,
~~~~k \in{\ZZ}_{n+1}\label{2.21b}
\bj
The last of the above equations agrees with the periodicity
property $K_k=K_{k+n+1}$ since ${\rm tr}F_i=0$. 
Following \cite{Bel}, one imposes the relations
\ai
&&\sum_{k\in {\ZZ}_{n+1}} \o^{r(1-k)}
\left( \frac{\rho_{kl}(\o^r\mu_{l})}
{\rho_{kl}(\mu_{l})}-\o^{-r} \frac{\rho_{k+1l}
(\o^r\mu_{l})}
{\rho_{k+1l}(\mu_{l})}\right)
\be_k(F_{l})=\delta_{r, r_{l}}
X_{l}Y_{l}\0\\
&&X_{l}=(1-\o^{r_l})
 \prod_{a\neq l} 
\frac{\o^{r_l}\mu_{l}-\mu_a}
{\mu_{l}-\mu_a}
\prod_a \frac{ \mu_{l}+\ep_{1a}}
{\o^{r_l}\mu_{l}+\ep_{1a}}\0\\
&&Y_{l}=\sum_{k\in {\ZZ}_{n+1}}
\left( 1+\mu_{l} \frac{d}{d \l}
\ln \frac{\rho_{kl-1}}{\rho_{k+1l-1}}
(\mu_{l+1})\right)\be(F_l)
\label{2.22a}
\b 
where
\a
\be_k(F_{l})=e^{K_k(F_{l})-\frac{f_{kl}}{2}}
\label{2.22b}
\b
In view of (\ref{2.21b}) one has 
\a
e^{-f_{kl}}=\frac{\be_k(F_{l})}{\be_{k-1}(F_{l})}
\label{2.22c}
\b
The functions $\rho_{al}~~,~~a \in {\ZZ}_{n+1}~~,~~~~l=1,...,N$
 are components of $(n+1)$-dimensional vectors
\a
{\bf \rho}_l(\l)=\D^{(l\,l-1)}(\l){\bf \rho}_{l-1}(\l)~~,~~~~~~
{\bf \rho}_{l-1}(\l)=\D^{(l-1\,l)}(\l){\bf \rho}_{l}(\l)
\label{2.22d}
\b 
where the entries of the matrices $\D^{(l\,l-1)}$
 and $D^{(l-1\,l)}$ are
\a
&&\D^{(l\,l-1)}_{ab}(\l)=\frac{\ga_l \be_b(F_l)}{(n+1)\rho_{bl-1}(\mu_l)}
\sum_{q\in {\ZZ}_{n+1}}\o^{q(a-b)} 
\frac{\l-\o^qe^{f_{bl}}\mu_l}{\l-\o^q\mu_l}\0\\
&&\D^{(l-1\,l)}_{ab}(\l)=\frac{\rho_{al-1}(\mu_l)}
{(n+1)\ga_l \be_b (F_l)}
\sum_{q\in {\ZZ}_{n+1}}\o^{q(a-b)} 
\frac{\l-\o^qe^{-f_{bl}}\mu_l}{\l-\o^q\mu_l}\0\\
&&\ga_l=\left( \prod_{p\in {\ZZ}_{n+1}}
\rho_{pl-1}(\mu_l))\right)^{\frac{1}{n+1}}\label{2.22e}
\b
The relation (\ref{2.22d}) together with 
\a
\rho_{j0}(\l)=\frac{1}{n+1}
\label{2.22f}
\b
determines uniquely $\rho_l~,~~l=1,...,N$. Finally the
entries of the diagonal matrices $P_l$ are fixed by
\a
e^{K_a(F_{l-1})+\frac{p_{al-1}}{2}}&=&\frac{ \rho_{al-1}(\mu_l)}{\ga_l}
\label{2.22g}
\bj 
Now we are in a position to formulate the 
following

${\bf Proposition \, 2}$ \cite{ Bel}: {\it Consider the $N$-soliton solution
 (\ref{2.10a}), (\ref{2.10b}). Denote by $g_+~~(g_-)$ the 
analytic continuation of (\ref{2.13a}) around 
$\l=0~,~~(\l= \infty)$. Then the following factorization 
is valid} 
\ai
g_{\pm}=g_{\pm}(N)g_{\pm}(N-1)...g_{\pm}(1),~~~~~
g_{\pm}(i)=e^{K(F_i)+P_i}e^{W_\pm(i)}
\label{2.23a}
\b
{\it where $K(F_i)$ and $P_i$ are fixed recursively by (\ref{2.22a})-(\ref{2.22f}) and $W_\pm(i)$ are the loop group elements}
\a
&&W_{\pm}(i)= \sum_{k\in{\ZZ}_{n+1}} f_{ki}W_{\pm}^k(\mu_i)\0\\
&&W_{+}^k(\mu)=-\sum_{r=1}^n\left(\frac{1-\o^{-rk}}{1-\o^{-r}}F_0^r
+\o^{r(1-k)}\sum_{p \geqslant 1}\frac{F_p^r}{\mu^p}\right)
\0\\
&&W_{-}^k(\mu)=\sum_{r=1}^n\left(-\frac{1-\o^{r(1-k)}}{1-\o^{-r}}F_0^r
+\o^{r(1-k)}\sum_{p \leqslant -1}\frac{F_p^r}{\mu^p}\right).
\label{2.23b}
\bj
Note that $W_{\pm}^k(\mu)$ do not contain contributions belonging
 to the principal Heisenberg subalgebra. Moreover 
\ai
&&W^k(\mu)=W_{+}^k(\mu_i)-W_{-}^k(\mu_i)=-\sum_{r=1}^n\o^{r(1-k)}F^r(\mu)\0\\
&&F^r(\mu)=\sum_{p \in {\ZZ}}\frac{F_p^r}{\mu_i^p}
\label{2.24a}
\b
and the element $F^r(\mu)$ diagonalize the adjoint action
 of the principal Heisenberg subalgebra:
\a
\left[\E_k, F^r(\mu)\right]=(\o^{rk}-1)\mu^kF^r(\mu)
\label{2.24b}
\bj
The Proposition 2 together with (\ref{2.24a})-(\ref{2.24b})
suggests a relation between the dressing symmetry and 
vertex operator representation of the soliton tau function
\cite{ jap, DT}.

\section{ Vertex operator representation for the soliton tau functions }

\setcounter{equation}{0}
\setcounter{footnote}{0}

The $A_n^{(1)}$ Toda models have a conformally invariant
extension \cite{ cat} known as the $A_n$ Conformal Affine 
Toda (CAT) model. The $A_n$ (CAT) equations are also integrables
 and admit zero curvature representation for the connection
 (\ref{2.2b})  in the affine Lie algebra $\hat{\G}=\hat{Sl}(n+1)=A_n^{(1)}$.
The affine algebra counterpart of $\Phi$ (\ref{2.2c}) is
\a
\Phi \rightarrow \Phi+\eta \hat{d}
+\frac{\hat{c}}{2(n+1)}\zeta
\label{3.1}
\b
where $\hat{d}$ and $\hat{c}$ stand for the derivation and the
 central charge respectively (c.f. (\ref{2.18a})). Combining
 (\ref{2.2a}), (\ref{2.2b}) and (\ref{3.1}) one obtains the 
$A_n$ CAT equations
\a
&&\del_+ \del_- \var_i =m^2 e^{2\eta}(e^{\var_i-\var_{i+1}}
-e^{\var_{i-1}-\var_i})\0\\
&&\del_+ \del_- \eta =0\0\\
&&\del_+ \del_- \zeta =m^2e^{2\eta}\sum_{i\in {{\ZZ}_{n+1}}}
e^{\var_i-\var_{i+1}}
\label{3.2}
\b
The first of the above equations coincides with the 
affine Toda equations (\ref{2.1}) for $\eta =0$.
According to \cite{BB}, affine solitons arise
after imposing the last restriction, it is also
worthwhile to recall the Hirota bilinear representation \cite{Hol}
\a
&&\del_+\tau_k\del_-\tau_k-\tau_k\del_+\del_-\tau_k=
m^2(\tau_{k+1}\tau_{k-1}-\tau_k^2)\0\\
& &e^{-\var_k}=\frac{\tau_k}{\tau_{k-1}}~~~,~~~~~
e^{\zeta_0-\zeta}=\prod_{k\in{\ZZ}_{n+1}}\tau_k
\,\,\,\,,k\in{\ZZ}_{n+1}\0\\
&&\zeta_0={(n+1)}m^2x^+x^-
\label{3.3}
\b
of the system (\ref{3.2}) for $\eta =0$. 
Setting $\zeta=\zeta_0$ with $\var_k=0~,~~k \in {\ZZ}_{n+1}$ 
and $\eta =0$ in (\ref{3.1}) one ends with the vacuum 
solution of the $A_n$ CAT equations (\ref{3.2})
\a
\Phi_0=\frac{m^2}{2}\hat{c}x^+x^-
\label{3.4}
\b
In \cite{ BB} it was argued that the Riemann 
problem (\ref{2.12}) is  related to a 
similar factorization problem in the affine
 group. More precisely, the following relation
\a
& &\widehat{g}_\pm(x)=e^{\pm\frac{\zeta_0-\zeta}{2(n+1)}\hat{c}}g_{\pm}(x)
\label{3.5}
\b
holds. In the above equation $g_{\pm}$ represent a
dressing transformation in the loop group which by
(\ref{2.11a}) creates $N$-soliton solutions 
from the vacuum; $\widehat{g}_\pm(x)$ stand for 
the affine group analog  of $g_{\pm}(x)$ (\ref{2.12}).

In what follows we shall need some basic facts concerning
the representation theory of the Lie algebras $A_n^{(1)}$
in the principal gradation \cite{ Kac}. We will be interested
on highest weight representations which are generated by the action of arbitrary
polynomials on the negative grade elements (\ref{2.18b})
on certain highest weight state $|\L>$. The later is annihilated
by the elements of positive grade
\ai
X_n|\L>=0~~,~~~~\left[\hat{d}, X_n\right]=nX_n,~~n>0
\label{3.6a}
\b
In the particular example of $A_n^{(1)}$, the highest
 weight state is characterized by (\ref{2.20a})
\a
&&F^r_0|\L>=\L(F^r_0)|\L>, \quad r=1,\ldots\, , n \0\\
&&\hat{d}|\L>=\L(\hat{d})|\L>, \quad
\hat{c}|\L>=\L(\hat{c})|\L>\0\\
\label{3.6b}
\bj
where $\L$ is a linear functional on the 
subalgebra $\hat{\G}_0$ of elements of $\ZZ$--grade  zero (cf. (\ref{2.18a})).

The principal Heisenberg subalgebra (\ref{2.20c}) admits 
a Fock representation. It is built up on the Fock vacuum
 $|0>$ 
\a
& &\E_k|0>=0,~~~~k \geq 1 \0\\
& &<0|\E_k=0,~~~~k \leq -1. \label{3.7}
\b 
When the value of $\hat{c}$ (\ref{2.18a}) is one, all the irreducible highest weight representations of the Lie algebra $A^{(1)}_n$ are expressed in terms 
of the (principal) Heisenberg subalgebra. The corresponding $A_n^{(1)}$ modules
are identified with the Fock space
 (\ref{3.7}). To be more precise, we introduce the "vertex
 operators": 
\a
&&V^r(\mu)={\rm exp}\{\sum_{k\leqslant -1}\frac{1-\o^{-kr}}{k}
~\frac{\E_k}{\mu^k}\}{\rm exp}\{\sum_{k\geqslant 1}\frac{1-\o^{-kr}}{k}
~\frac{\E_k}{\mu^k}\} \0\\
&&\hskip3cm r=1,...,n
\label{3.8}
\b
The following theorem is due to Kac

${\bf Theorem \,1}$ \cite{ Kac}: {\it Consider the Fock space representation 
(\ref{3.7}) of the principal Heisenberg subalgebra. For each 
$l \in {\ZZ}_{n+1}$ and $r \in {\ZZ}_{n+1}\backslash \, \{0\}$ define}
\ai
F^r(\mu)=\frac{\o^{rl}}{(n+1)(\o^r-1)}\, V^r (\mu)\label{3.9a}\b

{\it Then}

(i) (\ref{3.9a}) {\it admits a Laurent expansion}
\a
F^r(\mu)=\sum_{p\in {\ZZ}} \frac{F^r_p}{\mu^p} \label{3.9b}\b

(ii) {\it For each $l\in \ZZ_{n+1}$ the Laurent  coefficients 
$F_p^r \, , \, p \in\ZZ\,\, {\rm and}\,\, r\in {\ZZ}_{n+1} \backslash \, \{0\}$ together with the harmonic oscillators $F^0_k=\frac{1}{n+1}\E_k, \, k \in 
{\ZZ} \backslash (n+1)\ZZ$ obey the commutation relations
 (\ref{2.20b}).}

(iii) {\it For each $l \in {\ZZ}_{n+1}$, the Fock vacuum 
$|0>$ (\ref{3.7}) is a highest state $|\L_l>$ characterized by} 
\a
F^r_0|\L_l>=\frac{\o^{rl}}{(n+1)(\o^r-1)}\, |\L_l> \, , \, \,
\hat{c}|\L_l>=|\L_l>\ ,\, \, \hat{d}|\L_l>=0.
\label{3.9c}\bj

The meaning of the above theorem is that there are at least $n+1$ irreducible representations of the Lie algebra  $A^{(1)}_n$
which can be expressed in terms of bosonic oscillators. The value of the central charge is $\hat{c}=1$.
It was shown  in \cite{ Kac} that the representations with 
highest weight vector $|\L_l>$  (\ref{3.9c}), $l \in \ZZ_{(n+1)}$ 
exhaust the inequivalent fundamental representations of the affine algebra.

In what follows, it will be important to consider also the group
 theoretical tau functions 
\a
\tau_{\L}(\Phi)=<\L|e ^{-2\Phi}|\L>=e^{-2\L(\Phi)}
\label{3.10}
\b   
where $\Phi$ is given by (\ref{3.1}) and $|\L>$ is a
highest weight vector. Setting $\eta=0$ in (\ref{3.1})
 and taking into account (\ref{3.9c}), one gets a relation with
 the Hirota tau functions (\ref{3.3})
\a
\tau_{\L_k}(\Phi)=e^{-\frac{\zeta_0}{n+1}}\tau_k(\Phi)
\label{3.11}
\b 
Let us recall the following result \cite{ BB, Lez}
\a
\frac{\tau_{\L}(\Phi)}{\tau_{\L}(\Phi_0)} = 
<\L|\hat{g_-}^{-1}(x)\hat{g_+}(x)|\L>
\label{3.12} 
\b
where $\hat{g}_{\pm}(x)$ (\ref{3.5}) are the elements of the affine 
group which generate by dressing transformation a solution $\Phi$,
 starting from the vacuum  $\Phi_0$ (\ref{3.4}). Let us consider an
$N$-soliton solution  (\ref{2.10a}),  (\ref{2.10b}) of the $A^{(1)}_n$
Toda model. Then the Proposition 2, (\ref{3.5}) together
 with (\ref{3.12}) suggest us to study the following affine group 
element
\a
&&h(\al)=e^{-\al W_-(\phi)}e^{\al W_+(\phi)},\0\\
&&W_{\pm}(\Phi)=\sum_{k\in {\ZZ}_{n+1}}\var_k\,W_{\pm}^k (\mu),\0\\
&&\sum_{k\in {\ZZ}_{n+1}}\var_k=0
\label{3.13}
\b
where $W^k_{\pm}(\mu)$ are given by (\ref{2.23b}). As a corollary
of the Theorem $1$ one gets the

${\bf Proposition\, 3}$  {\it Consider an arbitrary fundamental representation of the Lie algebra $A_n^{(1)}$. Then the following commutation relations are valid}
\ai
&&\hskip-2cm W_-(\Phi)V^r(\mu)-V^r(\mu)W_+(\Phi)=\frac{1}{n+1}
\sum_{s=1}^n \frac{\o^{s(k+1)}}{\o^s-1} \hat{\var}_s V^{r+s}(\mu),
\label{3.14a}\b
{\it $k \in \ZZ_{n+1}$ labels the fundamental representations of 
$A^{(1)}_n$; $V^r(\mu)$ are the "vertex operators" introduced by (\ref{3.8}) 
and $\hat{\var}_s$, $s \in  \ZZ_{n+1}$ is the discrete Fourier transform 
of $\var_s$ ( $\sum_ s \var_s =0$)}
\a
&&\hat{\var}_s =\sum_k \o^{-ks}\var_k, \quad
\var_k=\frac{1}{n+1}\sum_s  \o^{ks}\hat{\var}_s
\label{3.14b}
\bj 

From the above Proposition it follows that the {\it affine} group element
(\ref{3.13}) admits the following (Taylor) expansion
\a
&&h(1)=e^{-W_-(\Phi)}\, e^{W_+(\Phi)}=\0\\
&&=\sum_{l=0}^{\infty}\frac{(-1)^l}{l!}
\sum_{s_1,...,s_l=1}^{n}\prod_{j=1}^{l}\Psi_{s_j}^{(k)}(\Phi)
V^{s_1+...+s_l}(\mu)\0\\
&&\Psi_s^{(k)}(\Phi) =\frac{1}{n+1}\frac{\o^{s(k+1)}}{\o^{s}-1}\widehat{\var}_s, \quad 
k\in \ZZ_{n+1} \label{3.15}
\b
 In view of the periodicity condition $V^r=V^{r+n+1}$ (cf. (\ref{3.8})), it is opportune to introduce the commutative associative algebra  $\F$ of 
(complex) dimension $n+1$. It is generated by $V^s$, $s\in \ZZ_{n+1}$.
The multiplication is 
\a
V^{r}\star V^{s}=V^{r+s}\label{3.16}
\b
It is worthwhile to note that the above equation describes "fusion rules" of a 
class of Rational Conformal Field Theories \cite{Ver}.  Introducing the invariant inner product 
\a 
<V^r,V^s>_{\F}=\delta^{(n+1)}_{r+s, 0}, \label{3.17}
\b
$\F$ becomes a Fr\"obenius algebra. Taking into account  (\ref{3.15}) and
(\ref{3.16}) one gets
 \a
h(1)=\F~{\rm exp}\{-\sum_{s=1}^{n}\Psi_s^{(k)}(\Phi)V^s\}(\mu)\label{3.18}
\b
where the symbol $\F$ means that the exponential is taken in the algebra 
(\ref{3.16}).

Diagonalizing the  "fusion rules" (\ref{3.16}) we get

${\bf Proposition \, \, 4}$ {\it The affine group element $h=h(1)$ (\ref{3.13})
is given by the expression}
\a
&&h=\frac{\hat{\be}_0(\Phi)}{n+1}+\sum_{p\in {\ZZ_{n+1}}}
(\o^p-1)\hat{\be}_p(\Phi) F^p(\mu)\label{3.19}
\b
{\it within any fundamental representation  (\ref{3.9a})--(\ref{3.9c}).
The numbers $\hat{\be}_p(\Phi)$ are the discrete Fourier transform 
(\ref{3.14b}) of $\be_k(\Phi)$ (\ref{2.21b}), (\ref{2.22b}).}

In \cite{Bel} we derived an exact expression for the adjoint action of the 
elements $g_{\pm}(i)$, $i=1, \ldots  ,\, N$  defined by (\ref{2.23a}),
(\ref{2.23b}) on the affine Lie algebra. The result is
\ai
&&g^{-1}_+(i)F^s(\o^c\l) g_+(i)=
\sum_{r,v \in {\ZZ}_{n+1}} R^{sc}_{rv}(i;\l)\left(
F^r(\o^v\l)-Z^{rv}[F_i](i;\l)\right)\0\\
&&\hskip4cm |\mu_i|\,>  \,|\l| \label{3.20a}
\b
where
\a
&&R^{sc}_{rv}(i;\l)=P^{c}_{v}(i;\l)\,Q^{s+c}_{r+v}(i;\l).
\label{3.20b}
\b
The matrices $P$, $Q$ and the vector $Z$ are expressed in terms of the 
quantities (\ref{2.22d}) and (\ref{2.22e}) as follows
\a
&&\hskip-0.5cm P^c_v(i;\l)=\frac{\ga_i}{n+1}\sum_{a,b \in {\ZZ}_{n+1}}
\o^{(a-1)v-(b-1)c}
\frac{1}{\rho_{a i-1}(\mu_i)} \D^{(i-1\, i)}_{ab} (\l)\rho_{bi}(\mu_{i+1})\0\\
&&\hskip-0.5cm Q^s_r(i;\l)=\frac{1}{(n+1) \ga_i}
\sum_{a,b \in {\ZZ}_{n+1}} \o^{(a-1)s-(b-1)r}
\frac{1}{\rho_{a i}(\mu_{i+1})} \D^{(i\, i-1)}_{ab} (\l)\rho_{bi-1}(\mu_{i})\0\\
&&\hskip-0.5cm Z^{sc}[F_i](i,\l)=\frac{\l\o^c}{\ga_i (n+1)^2 (\o^{s+c}\l-\mu_i)}
\sum_{a,b \in {\ZZ}_{n+1}} \o^{(s+c)a-bc} \D^{(i\, i-1)}_{ab}(\l)
\frac{\rho_{b i-1}(\mu_i)}{\be_a(F_i)}+\0\\
&&+\frac{\l \o^c \de^{(n+1)}_{s,0}}{(n+1)(\mu_i-\o^{s+c}\l)}
\label{3.20c}
\bj
As a consequence of (\ref{3.20b}), (\ref{3.20c}), (\ref{2.22d}) and taking into account the indentity
\a   
&&\hskip -1.1cm \frac{d\, \rho_{kj}}{d\l}(\l)=-\frac{1}{n+1}
\sum_{l=1}^j \sum_{a,a',r \in {\ZZ}_{n+1}}   
\frac{\o^{r(a'-a)}}{\l-\o^r\mu_l}\D^{(j\, l-1)}_{ka'}(\l)
\rho_{a'\,l-1}(\l)
\left( \frac{\rho_{al}(\l)}{\ga_l\be_a(F_l)}-\frac{\rho_{al-1}(\l)}
{\rho_{al-1}(\mu_l)}\right)\0
\b
one arrives at

${\bf Proposition \, 5}$ {\it Let $\mu_1, \ldots ,\, \mu_N$ are radially 
ordered complex numbers $|\mu_1|> \ldots ,\, > |\mu_N|$. Then the matrices
$R(j;\mu_i)$ ($j<i$) which act on $\CC^{n+1}\otimes \CC^{n+1}$ as well as the vectors $Z[F_j](j, \mu_i)\, \in \, \CC^{n+1}\otimes \CC^{n+1}$ ($j<i$) satisfy the equations}
\a
&&\hskip -2cm \sum_{p\in {\ZZ_{n+1}}} \o^{p(1-k)} \left(R(j;\mu_{j+1})\ldots R(1; \mu_{j+1})\right)^{p0}_{rv}=
\o^{r(1-k)} \de^{(n+1)}_{v,0} \frac{\rho_{kj}(\o^r\mu_{j+1})}{\rho_{kj}(\mu_{j+1})}\0\\
&&\hskip -2.5cm \sum_{\stackrel{1\leq l \leq j}{p\in {\ZZ_{n+1}}}}  
\o^{p(1-k)} \left(R(j;\mu_{j+1})\ldots R(l; \mu_{j+1})\,
Z[-F_l](l;\mu_{j+1})\right)^{p0}=
-\frac{\mu_{j+1}}{n+1}\, \frac{ d\ln \rho_{kj}}{d \l} (\mu_{j+1})\0\\
\label{3.21}
\b

Combining the above Propsition with Proposition 2 and the Proposition 4
we obtain the following 

${\bf Theorem \, 2}$ \cite{Bel} {\it Let $\Phi =\frac{1}{2}\sum_{k\in {\ZZ}_{n+1}}
\var_k |k><k|$ be an $N$--soliton solution of the $A_n^{(1)}$ Toda equations 
(\ref{2.10a}), (\ref{2.10b}), with $|\mu_1|> \ldots ,\, > |\mu_N|$. $g_{\pm}$
are supposed to be same as in the Proposition 2. Then for each fundamental representation of the Lie algebra $A_n^{(1)}$, the following identities are 
valid}
\a
&& {\rm Ad }\left( \widetilde{g}^{-1}_{+}(1) \ldots 
\widetilde{g}^{-1}_{+}(i-1)\right)\, \widetilde{g}(i)=
Y_i \left(1+ X_i F^{r_i}(\mu_i)\right)\label{3.22}
\b
{\it The factors $X_i,\, Y_i$ were introduced in (\ref{2.22a}).}

As a corollary of the above Theorem and taking into account (\ref{3.5}) and (\ref{3.12}) we get
\a
&&\frac{\tau_{\L}(\Phi)}{\tau_{\L}(\Phi_0)}=
e^{\frac{\z_0-\z}{n+1}}\prod_{i=1}^N Y_i \cdot
<\L|(1+X_1F^{r_1}(\mu_1))\ldots (1+ X_N F^{r_N}(\mu_N))|\L>\0\\
\label{3.23}
\b
Therefore (\ref{1.1a}) and (\ref{1.1b}) are equivalent provided that 
\a
&&\prod_{i=1}^N Y_i= {\rm exp}\left( \frac{\zeta-\zeta_0}{n+1}\right)
\label{3.24}
\b
Note that as a consequence of (\ref{3.3}), (\ref{3.11}), the last identity can be equivalently written as 
\a
&&\hskip-1.5cm \prod_{k\in{\ZZ}_{n+1}} \prod_{i=1}^N Y_i <\L|(1+X_1F^{r_1}(\mu_1))\ldots (1+ X_N F^{r_N}(\mu_N))|\L>=1\label{3.25}
\b
We hope to go back to the demonstration of the above identity in a future publication.

\section{ Remarks and conclusions }

\setcounter{equation}{0}
\setcounter{footnote}{0}

 The present talk is devoted to the study of the relation between the vertex operator representation of the (soliton) tau functions and the group of dressing transformations. We recall \cite{STS,fr} that the dressing group is dual to the corresponding loop (or affine) Lie group. On the other hand,  from the general analysis advanced by Leznov and Saveliev \cite{Lez}, it follows that
\ai
&&\tau_{\L}^g=<\L| {\rm exp}\{-mx^+\E_1\}\, g \, {\rm exp}\{mx^-\E_{-1}\} 
|\L> \label{4.1a}
\b
where $\E_{\pm 1}$ are the elements of $\ZZ$--grade $\pm 1$ of the Heisenberg subalgebra (\ref{2.20c}), $g$ is a constant element of the affine Lie group 
and $|\L>$ is a highest weight vector, is a tau function 
(\ref{3.10}) of the corresponding CAT equations. More precisely,
\a
&&{\rm exp}\{-2 \L (\Phi^g)\}=<\L| {\rm exp}\{-2\Phi^g\}|\L> 
\label{4.1b}
\bj
is a solution of the $A_n$ CAT equations (\ref{3.2}). On the other hand, it was observed in \cite{ DT} that setting in (\ref{4.1a})
\a
&&g={\rm exp}\{ \al_1 F^{r_1}(\mu_1)\}\ldots 
{\rm exp}\{ \al_N F^{r_N}(\mu_N)\}\label{4.2}
\b
where 
\a 
&&F^r(\mu)=\sum_{p\in {\ZZ}_{n+1}} \frac{F^r_p}{\mu^p}\0
\b
were introduced by (\ref{2.20b}), one recovers the $N$--soliton solutions of the CAT model. However, the relation with the dressing symmetry seems rather 
obscure. A hint of how to relate the group--algebraic approach \cite{DT,Lez}
with the group of dressing transformations has been suggested in \cite{BB}
on the particular example of the sine--Gordon equation.

Our approach generalizes the the results of 
\cite{BB} for arbitrary $A_n^{(1)}$ Toda models. However, some problems remain open. First, we have not considered the relation  with the B\"acklund transformations. We believe that the solution of this  problem 
can shed a new light on the geometry of affine Toda models. For arbitrary
simple Lie algebras one can generalize the Proposition 2, but it is not 
generally possible to obtain a matrix solution of the underlying linear 
problem. The reason is that the order of the automorphism (\ref{2.3a}), (\ref{2.3a})  is generically smaller than the dimension of any irreducible representation of the 
Lie algebra.


{\bf Acknowledgements} 

The authors are grateful to the organizers of the IV {\it International 
Conference on Non Associative Algebra and its Applications} for the
kind invitation to present this talk at the University of S\~ao Paulo, Brazil. 
We acknowledge L. A. Ferreira for discussions on the subject, and especially for explaining to us the results of \cite{Fer}. We thank G. Cuba for his collaboration on the early stage of \cite{Bel}. The work of H. B. and R. P. 
are supported financially by CNPq--Brazil and FAPERJ--Rio de Janeiro respectively. Both the research foundations are deeply acknowledged.

\end{document}